\documentclass{article}
\usepackage{arxiv}
\usepackage[utf8]{inputenc} 
\usepackage[T1]{fontenc}    
\usepackage{hyperref}       
\usepackage{url}            
\usepackage{booktabs}       
\usepackage{amsfonts}       
\usepackage{nicefrac}       
\usepackage{microtype}      
\usepackage{lipsum}
\usepackage{graphicx}
\usepackage{amssymb}
\usepackage{amsmath}
\usepackage{enumitem}
\usepackage{xcolor}
\usepackage{listings}  
\usepackage[numbers]{natbib} 

\title{The Logarithmic Relationship between Mean Arterial Pressure and Heart Rate}

\author{
 Maximilian M. Nguyen$^1$\\
  School of Medicine, Emory University$^1$\\
  110 Woodruff Circle, Atlanta, GA 30322 \\
  \texttt{maximilian.nguyen@emory.edu}
}

\begin{document}
\maketitle
\pagestyle{plain}

\begin{abstract}
The classic heuristic formula for calculating the mean arterial blood pressure (MAP) in the human cardiac cycle using $\frac{1}{3}$ of the systolic pressure ($P_S$) plus $\frac{2}{3}$ of the diastolic pressure ($P_D$) importantly neglects the nonlinear effect of heart rate ($\text{HR}$) on blood pressure. Regression analysis on blood pressure data from patients experiencing a variety of heart rates captured a logarithmic relationship between mean arterial pressure and heart rate, $F_S=0.086 \cdot\text{ln}(\text{HR}-41.2)+0.079$, where $F_S$ is the MAP form factor corresponding to the fraction of time the cardiac cycle spends in ventricular systole. The logarithmic relationship is qualitatively distinct from the linear and exponential relationships previously suggested in the literature yet better aligns with the known human physiology. This heart rate-dependent formula for the mean arterial pressure recapitulates the classic intuition as a special case ($\text{HR}=60\text{ bpm} \implies F_S = \frac{1}{3}$). In addition, the model suggests several other MAP heuristics: $\text{HR}=45\text{ bpm} \implies F_S = \frac{1}{5}$, $\text{HR}=50\text{ bpm} \implies F_S = \frac{1}{4}$, $\text{HR}=175\text{ bpm} \implies F_S = \frac{1}{2}$. A calculator for the MAP can be found at \href{https://mngu265.shinyapps.io/map\_calculator/}{\text{https://mngu265.shinyapps.io/map\_calculator/}}.
\end{abstract}



\maketitle
\section*{Introduction}
The mean arterial blood pressure (MAP) is a key measure of hemodynamics during the cardiac cycle. It is defined as the average arterial blood pressure over both the systolic and diastolic phases of a cardiac cycle and is an important proxy for the systemic resistance in the peripheral vasculature. Clinicians commonly incorporate MAP as a factor in determining their treatment and intervention plans. Studies have suggested that MAP values can be linked to both the pathophysiology and treatment of various conditions across the gamut of medicine \cite{sesso_systolic_2000, l_clinical_1989, cnossen_accuracy_2008, hawryluk_mean_2015, leone_optimizing_2015, haque_analysis_2007}. 

Given the MAP's clinical significance, it is important to get an accurate estimation of its value. There are several types of approaches to calculate it. The most accurate method is through a time-averaged integration of the blood pressure waveform over the cardiac cycle on direct arterial measurements obtained by catheter-manometer systems. However, this is necessarily the most invasive approach as it requires catheterization. A more non-invasive approach is integration of the blood pressure waveform from measurements taken via automated oscillometric sphygmomanometers \cite{orourke_pulse_2001}. The integration calculation in such devices is typically proprietary. 

Before the advent of modern electronic measurement devices, practitioners relied on mathematical formulas to estimate the MAP. Today, these formulas are used as a simple heuristic for clinicians to make an initial MAP determination. The waveform of a typical cardiac cycle is complex, making it difficult to parameterize with elementary functions. To make the calculation more tractable, formulas have typically assumed the MAP to be a simple time-weighted average of the systolic ($P_S$) and diastolic pressures ($P_D$), neglecting the transient dynamics of pressure in each stage of the cardiac cycle. Thus, MAP estimation formulas usually take the form:
\begin{equation}
    \text{MAP} = F_S P_S+F_DP_D
\end{equation}
where $F_S$ and $F_D$ represent the fraction of time the cardiac cycle spends in systole and diastole respectively, and $P_S$ and $P_D$ are the systolic and diastolic pressure respectively. This equation can be simplified by recognizing that if one is not in the systolic phase of the cardiac cycle, one is necessarily in the diastolic phase, i.e. $F_S+F_D=1$, which implies that $\text{MAP} = F_S P_S+(1-F_S)P_D$. Thus, there is only one parameter in this formula ($F_S$), which the literature typically refers to as the MAP form factor. Rearranging this equation and defining the pulse pressure ($\text{PP}$) as the difference between the systolic and diastolic pressures (i.e. $\text{PP}\equiv P_S-P_D$), we arrive at the MAP formula in an alternative form:
\begin{equation}
    \text{MAP}=F_S\cdot\text{PP}+P_D
\end{equation}

The first formula for the MAP dates back to the 1939 formula of Wezler and Böger \cite{wezler_dynamik_1939}, $F_S=0.42$. The most widely used heuristic, mainly due to its simplicity, is the Gauer formula \cite{gauer_kreislauf_1960} from 1960, $F_S = \frac{1}{3}$. The formula of Meaney et al. \cite{meaney_formula_2000} has also received some attention, $F_S=0.412$. However, all of these formulas are too simplistic in that they neglect the effect of heart rate ($\text{HR}$) on the MAP \cite{segers_amplification_2009}. Given that the human heart can operate across a wide range of heart rates under normal and pathological conditions (from less than 40 to in excess of 200 bpm), it would be inaccurate to assume that MAP is independent of heart rate. The first model in the literature for a heart rate-dependent MAP is the exponential model of Moran et al. \cite{moran_calculation_1995}, $F_S=0.01e^{4.14-\frac{40.74}{\text{HR}}}$. MAP models with a simple linear dependence on HR have also been proposed by Razminia et al. \cite{razminia_validation_2004}, $F_S=\frac{1}{3}+0.0012\text{HR}$. 

In this Letter, we reanalyze existing heart rate-dependent MAP data and show that the relationship between heart rate and MAP is best described by a logarithm. This improved model provides several new heuristics to quickly estimate the MAP at different heart rates. 

\section*{Results}
\begin{figure*}[!t]%
\centering
\includegraphics[scale=0.25]{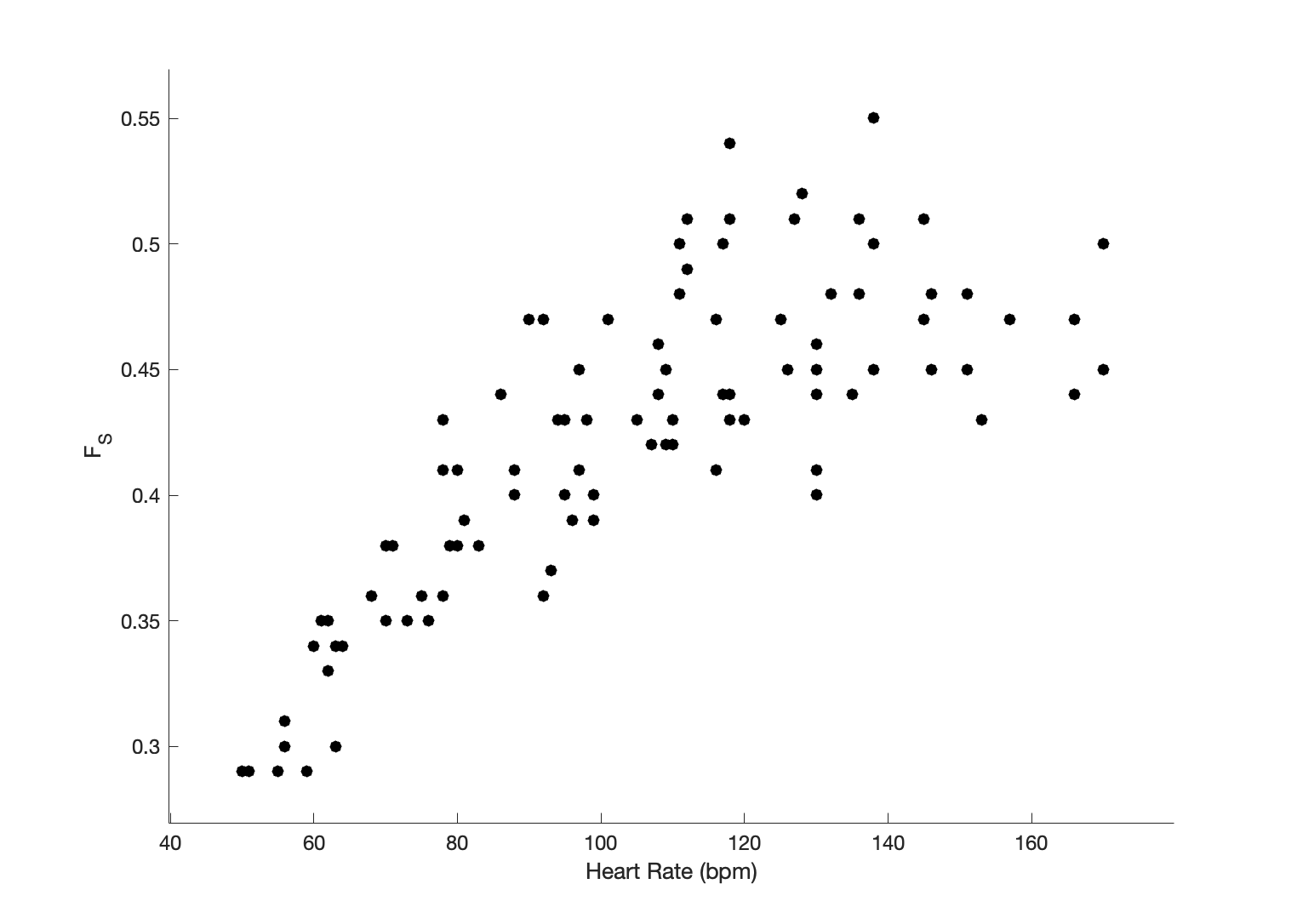}
\caption{MAP form factor ($F_S$) as a function of heart rate in a population of twenty males during exercise. Figure and data adapted from Moran et al. \cite{moran_calculation_1995}.}\label{fig:Fig1}
\end{figure*}

Data for this Letter is obtained directly from a study by Moran et al.\cite{moran_calculation_1995}. Briefly, the authors of that study obtained the heart rates and the durations of systole and diastole in twenty healthy male subjects exercising on a cycle ergometer at different work loads using sphygmomanometer and ECG recordings. From the ECG recordings, the researchers obtained the fraction of time the cardiac cycle spends in systole (i.e. $F_S$) from the measured R-R intervals. The corresponding heart rate to the ECG measurement was recorded. Measurements for all twenty subjects were then aggregated (Figure \ref{fig:Fig1}). 

\begin{figure*}[!t]%
\centering
\includegraphics[scale=0.15]{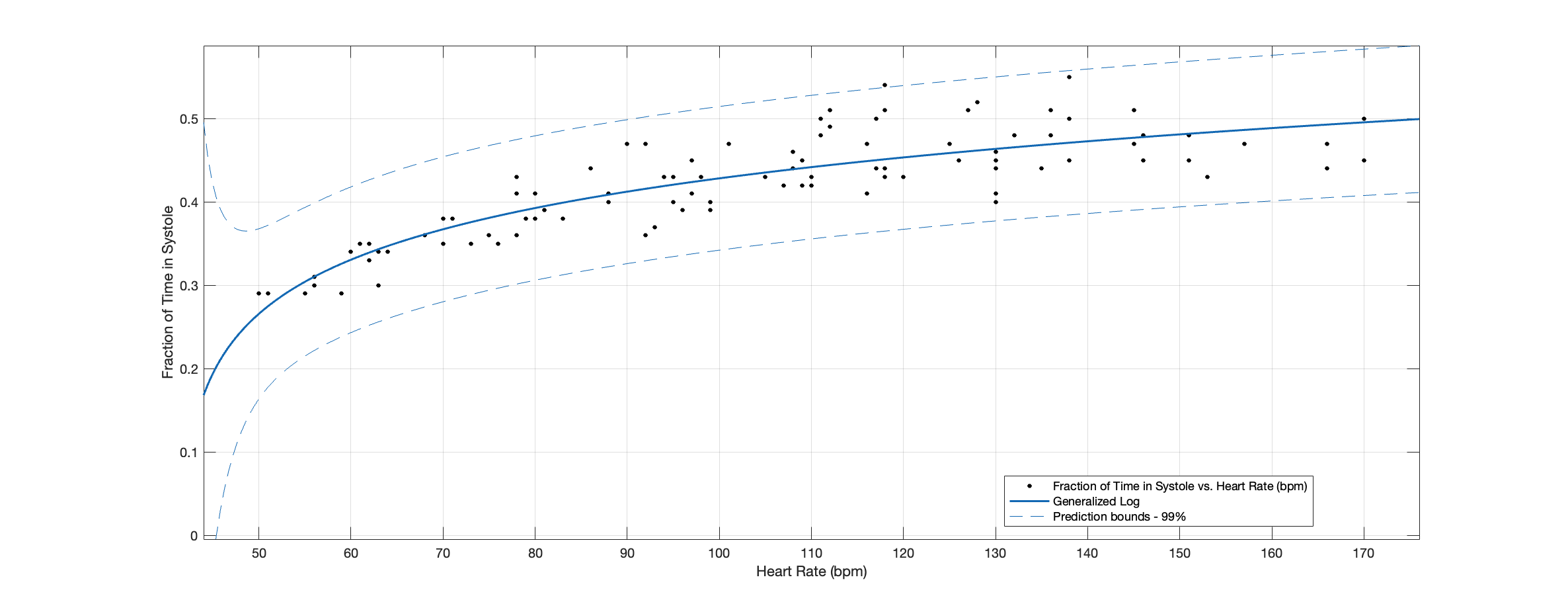}
\caption{MAP form factor ($F_S$) as a function of heart rate in a population of twenty males during exercise. Figure and data adapted from Moran et al. \cite{moran_calculation_1995}. Data fit to a generalized logarithmic model (blue curve given by $F_S=0.086 \cdot\text{ln}(\text{HR}-41.2)+0.079$). Bounds on model confidence intervals (99\%) given by dashed blue lines.}\label{fig:Fig2}
\end{figure*}

After performing a regression analysis of the data, we find that the most appropriate model for the relationship between the heart rate and the MAP form factor is an expression in the form of a generalized logarithm, i.e. $F_S=a \text{ ln}(\text{HR}-b)+c$ (Figure \ref{fig:Fig2}).
Fitting this data set produces the following model for the heart rate-dependent MAP form factor:
\begin{equation}
   F_S=0.086 \cdot\text{ln}(\text{HR}-41.2)+0.079
\end{equation}
While we expect the relationship to remain logarithmic, more data in the future would allow for a more robust estimation of the logarithm parameters. This model stands in contrast to the conclusion of Moran et al. \cite{moran_calculation_1995}, which suggested an exponential relationship. The reason a logarithm is a more appropriate fit from a mathematical standpoint can be understood by recognizing that the data displays a monotonically positive correlation that is concave down between $F_S$ and $\text{HR}$, which can be fit naturally with functions such as a logarithm and root functions. In contrast, an exponentially growing function cannot be concave down, and thus the Moran model is inconsistent with the data. In fact, the regression analysis shows an exponential model is a worse fit than simply assuming a linear relationship. A comparison of the data regression with different functions is shown in Table \ref{table:1}. 

\begin{table}[h!] 
\centering 
\begin{tabular}{||c c c c||} 
 \hline
 Model & Form & Adjusted $R^2$ & RMSE \\
 \hline\hline
 Logarithmic & $a \text{ ln}(\text{HR}-b)+c$ & \textbf{0.722} & \textbf{0.0325} \\ 
 \hline
 Root function (Power) & $a(\text{HR})^b$ & 0.669 & 0.0355 \\
 \hline
 Linear & $a \text{HR}+b$ & 0.618 & 0.0381\\
 \hline
 Exponential & $a e^{b\text{HR}}$ & 0.583 & 0.0399 \\
 \hline
\end{tabular}
\caption{Model Comparison for Relationship between Heart Rate and $F_S$}
\label{table:1}
\end{table}











\section*{Discussion}
While the gold standard in terms of accuracy to determine MAP values would be the integration of the blood pressure waveform obtained from arterial catheterization, a previous survey comparison of methods suggested that MAP estimations using formulas can produce reasonably accurate results \cite{papaioannou_mean_2016}. Importantly, estimation using formulas can be done quickly and non-invasively. Historically however, formulas have typically neglected the complex dependence of heart rate on the MAP. Here we obtain a formula, $F_S=0.086 \cdot\text{ln}(\text{HR}-41.2)+0.079$, that provides a more qualitatively accurate physiological relationship between heart rate and MAP form factor than has been previously described. Our model for a heart rate-dependent MAP is:
\begin{equation}
    \text{MAP}=[0.086 \cdot\text{ln}(\text{HR}-41.2)+0.079] \cdot \text{PP}+P_D
\end{equation}
This relationship features a logarithmic relationship between heart rate and MAP, which is a significant improvement over the previous suggestions of either a linear or an exponential relationship \cite{moran_calculation_1995, razminia_validation_2004}. In addition to the relationship being more aptly described mathematically by a logarithmic model, from a physiological standpoint, the choice is also more consistent. Firstly, we know that living humans have a lower limit on their heart rate, as low as 30's and 40's in conditioned athletes \cite{doyen_asymptomatic_2019}. Logarithmic functions naturally have a finite lower cutoff as an intrinsic feature. In contrast, exponentially growing and linear functions do not, making them less physiologically realistic. Secondly, while the duration of systole does shrink with increased heart rate \cite{moran_calculation_1995}, due to the physical limitations of the electrophysiology, we know that the duration of systole cannot become arbitrarily short \cite{harley_pressure-flow_1969}. This tapering feature of time spent in systole at larger heart rates is more appropriately accommodated by a logarithmic function, as opposed to either an exponential or linear function.

For the practitioner, this logarithmic model gives rise to new heuristics that can be used to quickly estimate a heart-rate dependent MAP. A list of MAP form factors, $F_S$ (i.e. the fraction of the cardiac cycle spent in systole), that are easier to remember is given in Table \ref{table:2}.

\begin{table}[h!] 
\centering 
\begin{tabular}{||c | c||} 
 \hline
 Heart Rate [$\text{HR}$] (bpm) & Fraction of Cardiac Cycle Spent in Systole ($F_S$)\\ [0.5ex] 
 \hline\hline
 45 & 1/5 \\ 
 \hline
 50 & 1/4 \\ 
 \hline
 60 & 1/3 \\ 
 \hline
 175 & 1/2 \\ 
 \hline 
 & \\
 \hline
 70 & 0.37 \\ 
 \hline
 80 & 0.39 \\ 
 \hline
 90 & 0.41 \\ 
 \hline
 100 & 0.43 \\ 
 \hline
 120 & 0.45 \\  
 \hline
\end{tabular}
\caption{Heuristics for Relationship between Heart Rate and $F_S$}
\label{table:2}
\end{table}
As we can see from the table, we recover Gauer's \cite{gauer_kreislauf_1960} commonly-used heuristic of systole being 1/3 of the cardiac cycle time as a special case. This model suggests the 1/3 heuristic is most appropriate to use when the patient has a heart rate of 60 bpm. When the heart rate is reduced to 45 bpm, the cardiac cycle only spends 1/5 of the time in systole. When the heart rate is 50 bpm, the cardiac cycle spends 1/4 of the time in systole. And in the other direction, when the heart rate increases to the neighborhood of 175 bpm, the cardiac cycle spends 1/2 the time in systole. Due to the logarithmic relationship, MAP is most sensitive to the effect of heart rate at low heart rates and less sensitive at higher rates. These heuristics can be used to quickly calculate the MAP if one knows the heart rate, systolic, and diastolic pressures. 

We have made a heart rate-dependent MAP calculator for users, which can be accessed at \href{https://mngu265.shinyapps.io/map\_calculator/}{\text{https://mngu265.shinyapps.io/map\_calculator/}}.

As this model was derived from the Moran et al. dataset \cite{moran_calculation_1995}, which only had twenty participants, future work should aim to expand the dataset to make the model more robust, incorporating more participants of different gender, age, fitness, and socioeconomic backgrounds. While we expect the logarithmic paradigm to generalize, the parameters of the model may need to be revised as new data is acquired. Increasing the number of participants should also improve the heteroskedasticity seen in the present dataset, which would improve the accuracy of the MAP heuristics.

\section*{Methods}
The data was analyzed in MATLAB. Regression analysis was performed using the `curve fitter' toolbox. The user application to calculate MAP was created in RStudio and is hosted by ShinyApps.

\section*{Data Availability}
All data produced in the present work are contained in the manuscript.

\section*{Code Availability}
Code is available upon request.


\section*{Author Contributions}
M.M.N. designed research, performed research, and wrote and reviewed the manuscript.

\section*{Competing Interests}
The author declares no competing financial or non-financial interests.

\newpage
\bibliography{MAP}

\section*{Figures and Tables Legend}
Figure 1. \textit{MAP form factor ($F_S$) as a function of heart rate in a population of twenty males during exercise. Figure and data adapted from Moran et al. \cite{moran_calculation_1995}.}
\\
Figure 2. \textit{MAP form factor ($F_S$) as a function of heart rate in a population of twenty males during exercise. Figure and data adapted from Moran et al. \cite{moran_calculation_1995}. Data fit to a generalized logarithmic model (blue curve given by $F_S=0.086 \cdot\text{ln}(\text{HR}-41.2)+0.079$). Bounds on model confidence intervals (99\%) given by dashed blue lines.}
\\
Table 1. \textit{Model Comparison for Relationship between Heart Rate and $F_S$.}
\\
Table 2. \textit{Heuristics for Relationship between Heart Rate and $F_S$.}
\end{document}